\documentclass[a4paper,11pt,pointlessnumbers]{scrartcl}
\usepackage{amsmath}
\usepackage{amsfonts}
\usepackage{amssymb,mathrsfs,wasysym}
\usepackage{t1enc}
\usepackage[latin1]{inputenc}
\usepackage[greek.ancient,english]{babel}
\usepackage{natbib}
\usepackage{amssymb}
\usepackage[final]{pdfpages}
\usepackage{caption}
\usepackage{url}

\usepackage[colorlinks=false, linktocpage=true,backref,pagebackref,hidelinks]{hyperref}

\newcommand{\beq}{\begin{equation}}
\newcommand{\eeq}{\end{equation}}
\newcommand{\beqarr}{\begin{eqnarray}}
\newcommand{\eeqarr}{\end{eqnarray}}
\newcommand{\beqa}{\begin{eqnarray*}}
\newcommand{\eeqa}{\end{eqnarray*}}
\begin{document}
\thispagestyle{empty}

\title{\bf \large From heliocentrism to epicycles: A commentary on 
pre-Ptolemaic astronomy}
\author{\normalsize Erhard Scholz\footnote{University of Wuppertal, Faculty of  Math./Natural Sciences, and Interdisciplinary Centre for History and Philosophy of Science, \quad  scholz@math.uni-wuppertal.de} }
 \date{\small \today} 
\maketitle
\begin{abstract}
If one wants to translate the  heliocentric  picture of planets moving uniformly on circular orbits about the sun  to the perspective of a terrestrial observer, using classical (ancient) geometric means only, one is naturally led to the investigation of epicyclic constructions. The  announcement of the heliocentric hypothesis by Aristarchos of Samos and the 
  invention  of the method of epicycles    happened during the 3rd and 2nd centuries BC. The latter developed   into  the central tool  of Hellenistic and Ptolemaic  astronomy. In the present literature on the history of astronomy the parallel rise of the heliocentric view and the methods of epicycles is   usually considered  as a pure contingency. Here I explain why I do not find this view convincing. 
\end{abstract}

\setcounter{tocdepth}{2}
\tableofcontents
\newpage
\section*{Introduction}
\addcontentsline{toc}{section}{\protect\numberline{}Introduction}
Often  epicyclic models of planetary motion are  considered as   contrary to the heliocentric view.
 This is, however,  wrong. If one wants to get an impression of how heliocentric heavenly motions   appear to a terrestrial observer, i.e, if one wants to picture the  phenomena  which follow from a heliocentric ontological assumption, one is naturally lead  to the construction of  nested cyclical  motions by a kind of {\em kinematic inversion}. The underlying transformation can easily be developed 
 in the framework of   ancient Euclidean geometry; it  leads without fail to a set of graphical methods using cycles and epicycles.

Historical evidence for the transition from heliocentrism to epicyclic kinematics is sparse. This seems to be the reason for the deplorable fact that an interdependence of heliocentrism and epicyclic kinematic models is seldom discussed in the literature. Often a connection is  even rejected, although with insufficient arguments as will be discussed in section \ref{section discussion} of this paper. 
In my view this is a surprising and deplorable gap in the interpretation of the sources of Greek astronomy. Although I am neither a historian of astronomy, nor could I present new sources on the origin of the epicyclic approach,  the preparation of the present book for Jesper L\"utzen offers  a welcome  opportunity to me to develop the simple argument indicated above in written form.\footnote{I presented the argument in several talks years ago; among them 2013 at a conference in Aarhus. Jesper made me aware of Aaboe's appreciation of the kinematic inversion (without using the word) at the Aarhus conference in 2013. \label{fn Aarhus}} 

	In chap. 2 of  his beautiful  exposition of Greek astronomy Asger Aaboe explains  the epicyclic models by a discussion of  the kinematic inversion first of the sun's   motion and then for  the planets. He emphasizes the importance of this discussion by the following remark:
\begin{quote}
For a proper appreciation of the Greek planetary models [the epicyclic ones, ES]  it is important to recognize that they are more than a mere collection of ad-hoc devices that can reproduce the planets' apparent behavior. In fact, when appropriately scaled, they turn out to be correct representations of the planets' motions relative to the earth in {\em distance} as well as in {\em direction}.  \citep[p. 82f.. emphasis in the original]{Aaboe:Episodes2} 
\end{quote} 
 Aaboe's stops short, however, of drawing conclusions from this observation for the  history of the transition from  heliocentrism to the method of epicycles. His  exposition in this section remains purely didactical; as such it makes good reading in particular for non-experts in ancient astronomical thought. 

As the present book is  written for broader audience, not particularly for historians  of astronomy, it may  be useful to remind  us of the rough time order of events in Greek astronomy insofar they concern our question by the following timetable. 

\subsection*{\small Timetable of Greek astronomy \label{subsection timetable}}
\fbox{\parbox{15cm}{
\begin{itemize}
\item  {\em Classical Greek astronomy}, {Description of heavenly motions by}  {\em homocentric spheres} with the earth at the center   (4th century BC):\\
	Eudoxos (ca 390 -- ca. 340 BC), Kallippos ( ca. 370-- 300 BC),\\
	Aristotle (384 -- 322 BC), \;  main source {\em De caelo}.
\item   Upsurge of a {\em heliocentric view} as a  counter-perspective (3rd and 2nd centuries BC): \\
	Aristarchos of Samos (310 -- 230 BC) \; (only indirect sources), \\
	Archimedes ($\sim$ 287 -- 212 BC),  \;  {\em Sand reckoner},\\
		Seleukos of Seleukeia ($\sim$ 180 -- $\sim$ 120 BC)\; 
(report by Plutarchos, 46 -- 120 AD).
 \item  Rise of ideas for an   {\em epicyclic kinematics} (3rd and 2nd centuries BC):\\
 Origin unclear, 
 usual picture in the literature: \\
 rise out of a sudden, about the mid 3rd c. BC. \\
Investigations  of epicycles by Apollonios  ($\sim$ 260 -- $\sim$ 190  BC)\;  (source: {\em Almagest}). 
Late 3rd and 2nd c. BC ff.  tool kit for epicyclic planetary models, \\
 competitive with homocentric spheres. 
 \item    Diverse planetary hypotheses (2nd century BC), \\
  critique  by {\em Hipparchos} of Nicaea ($\sim$  190  -- $\sim$ 120 BC)  (source: {\em Almagest}).\\
  During the following centuries cumulative development of geometric tools for epicyclic models in order to  {assimilate diverse ``anomalies''}\\
  in particular for the
  \begin{itemize}
	\item  sun  (old), 
	\item moon model (Hipparchos),
	\item   planets with  {two anomalies}  (source {\em Almagest}).
\end{itemize}
  Solution finally: decentered epicycles with {\em equant} point.
 \item  {Ptolemy} ($\sim 90$ -- $\sim$ 168 AD), full epicyclic model in the {\em Almagest}.\\
Fusion of epicyclic models with homocentric sphere view in {\em Planetary Hypotheses}. \\
Heliocentrism  no longer explicitly   mentioned.
\end{itemize}
Note:  Aristarchos'  {heliocentrism} and the first documented studies of  epicyclic model\\ by Apollonios are  at most one generation apart, both around mid 3rd. century BC.\\
}}
\nocite{Goldstein:1967}
\vspace{1.5em}

The  first part of following the paper explains the geometry of the kinematic inversion from a heliocentric to a geocentric view  in a systematic perspective, with side remarks to the historical context. For didactical reasons it  proceeds stepwise:  at first the inversion is considered for the sun (sec. \ref{subsection earth motion}), then for the interior planets Mercury,  Venus (sec. \ref{subsection inferior planets}) and finally for the outer planets Mars, Jupiter Saturn (sec. \ref{subsection superior planets qualitative}). Readers with a good imagination of kinematic constellations can skip these explanations, because they will grasp immediately how the kinematic inversion of a heliocentric picture of planetary motion on circular orbits to the perspective of a terrestrial observer leads to simple epicyclic constructions.  The last two subsections of the first part, 
however,  may be instructive also for them. Sec. \ref{subsection superior planets quantitative} gives a general survey of how observational data on the motion of the planets, available to Greek astronomers in the 3rd century BC, may be used to adapt the free parameters of these epicyclic constructions and, with it, the underlying heliocentric mechanisms. Although the theory of the moon stood in the center of many of the ancient Greek astronomical investigations, it will not appear in this paper, because its special status  among the ancient ``planets'' makes it less instructive for the differences between a heliocentric and  geocentric approach.

The second part of the paper (sec. \ref{section hist hypothesis}) turns towards the historical perspective. It explains why I do not consider it admissible to  neglect  the  historical relevance of the geometric argument developed  in sec. \ref{section kinematic inversion} for the development of  pre-Ptolemaic Greek astronomy. It is well known that the method of epicyclic motions  in Hellenistic Greece originated, and had its first phase of development, in the   roughly one and a half centuries between the life of the two main  protagonists of the heliocentric view, Aristarchos of Samos (fl. $\sim$  280 BC) and Seleukos of Seleukia (f. $\sim$ 150 BC), with Apollonios of Perga (fl. $\sim$ 230 BC) being the best known contributor to the theoretical study of epicyclic constellations just in the middle of the period. In my view  this cannot be downplayed as a pure coincidence. It rather seems extremely likely  that the authors on epicyclic motions of the 3rd and 2nd century BC knew of the relevance of their investigations for the heliocentric picture, and at least some of them were motivated by it. In particular the widely supported claim that neither Aristarchos nor Seleukos made any reasonable attempts for adapting the parameters of their heliocentric geometric pictures to observational data  does not seem particularly convincing in the light of  what we know on data availability to Greek astronomers in the 3rd and 2nd centuries BC  and of what have seen in sec. \ref{subsection superior planets quantitative}. 

Of course the beautiful, but too  simple assumption of uniform circular movement was an obstacle for adapting kinematically inverted epicyclic models of  heliocentric planetary motions to the data. The ensuing separation of a pragmatic use of epicyclic models from its heliocentric background is shortly discussed in sec. \ref{subsection Ptolemy}. 
We  learn  from Ptolemy's later report in the {\em Almagest} and Neugebauer's magisterial work on ancient astronomy \citep{Neugebauer:HAMA} that the difficulties of designing quantitatively satisfactory epicyclic models of planetary motions increased  about  the  middle of the 2nd century BC and  culminated in  a critical review by Hipparchos of Nicea, the towering contemporary of Seleukos. Hipparchos, also an impressive observational astronomer, seems to have been a sceptic with regard to theoretical speculations which did not lead to quantitative models living up to the latest precision data of the time. This may have contributed to sever  the  pragmatic treatment of epicyclic models from  its former link to the  heliocentric view even further. At the time of Ptolemy (fl. $\sim$ 120 AD) the link was close to forgotten, perhaps even consciously suppressed because of ideological   reasons. 

In the final discussion (sec. \ref{section discussion})  it will be explained why I find the present neglect of the role of heliocentric considerations on the trajectory of pre-Ptolemaian astronomy neither convincing nor fruitful.

\section{\small   Epicyclic models from kinematic inversion (systematic introduction) \label{section kinematic inversion}}

\subsection{\small Kinematic inversion of the earth's motion about the sun \label{subsection earth motion}}

  Aristarchos and other supporters of the heliocentric view had to address the question, {how  heliocentric motions} (the imputed ``true'' ones)  express themselves as ``apparent'' ones for a  terrestrial observer. 
 Until the times of Hipparchos to whom the first chord tables are attributed  heliocentric thinkers  had  to rely  mainly on    {geometrical methods}, later called ``synthetic'' in contrast to  trigonometric methods (or even algebraic/analytic ones) which were not  available at the time. 
 Lacking detailed historical sources,  we still can  look for appropriate geometrical constructions that may be used for converting  heliocentric motions into those seen by a terrestrial observer. In this section this will be done in a systematic way; in  sec. \ref{section hist hypothesis} it will then be discussed why this matters for our historical  understanding of Hellenistic astronomy. 
 
\begin{figure}[h]	
\hspace*{3em}	\includegraphics[angle=0,scale=0.7]{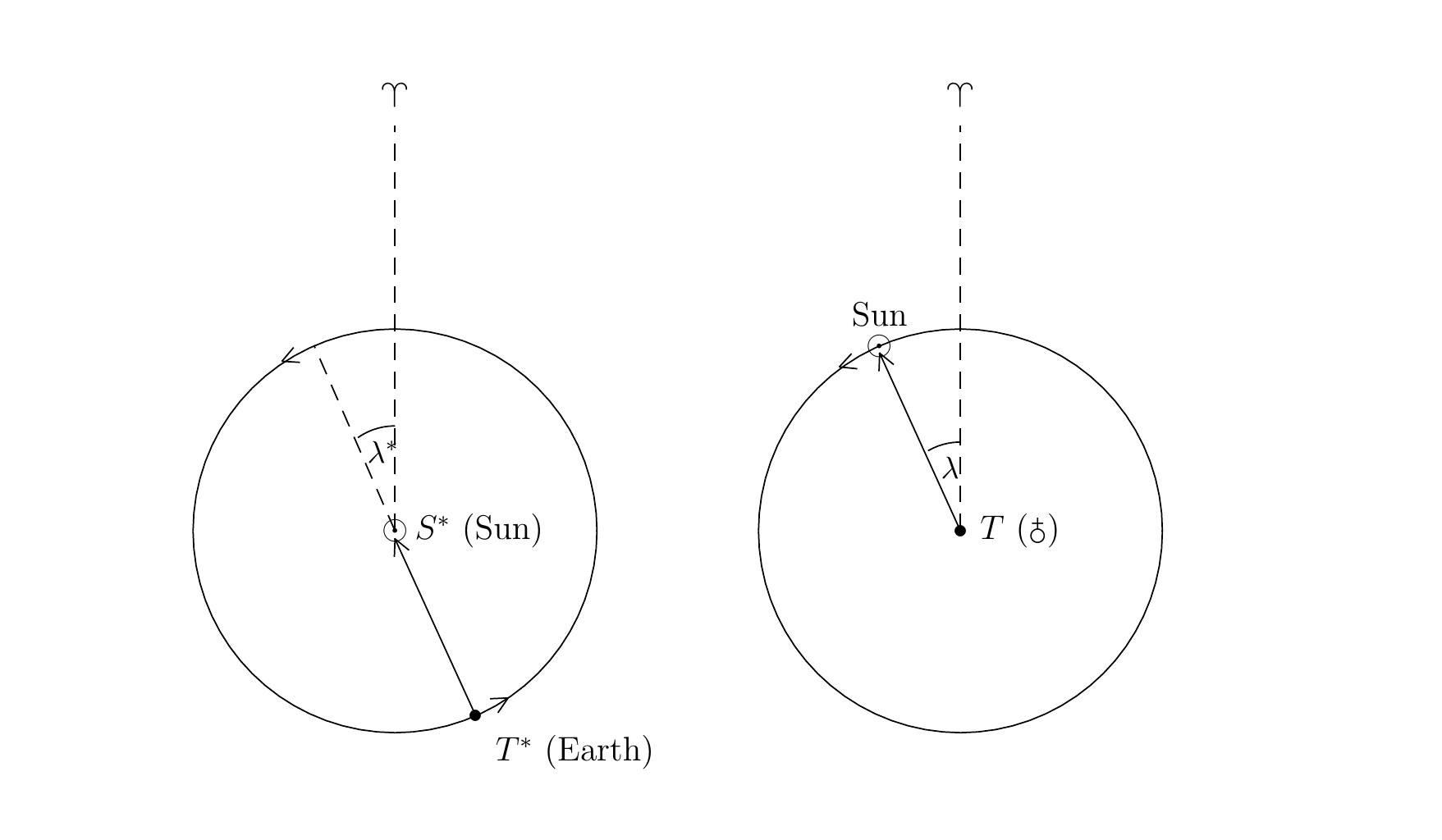} 
\caption{Kinematic inversion for the  motion of earth $\earth$ and sun $\astrosun$. Left: heliocentric  motion of  $T^{\ast}$ about    $S^{\ast}$. Right: inverted motion as seen from an observer at $T$. \\   $\lambda(t)$ time-dependent ecliptical longitude with respect to direction of aries $\aries$. \label{fig S-T}}
\end{figure}

\begin{figure}[h]
\hspace*{8em}\includegraphics[scale=0.6]{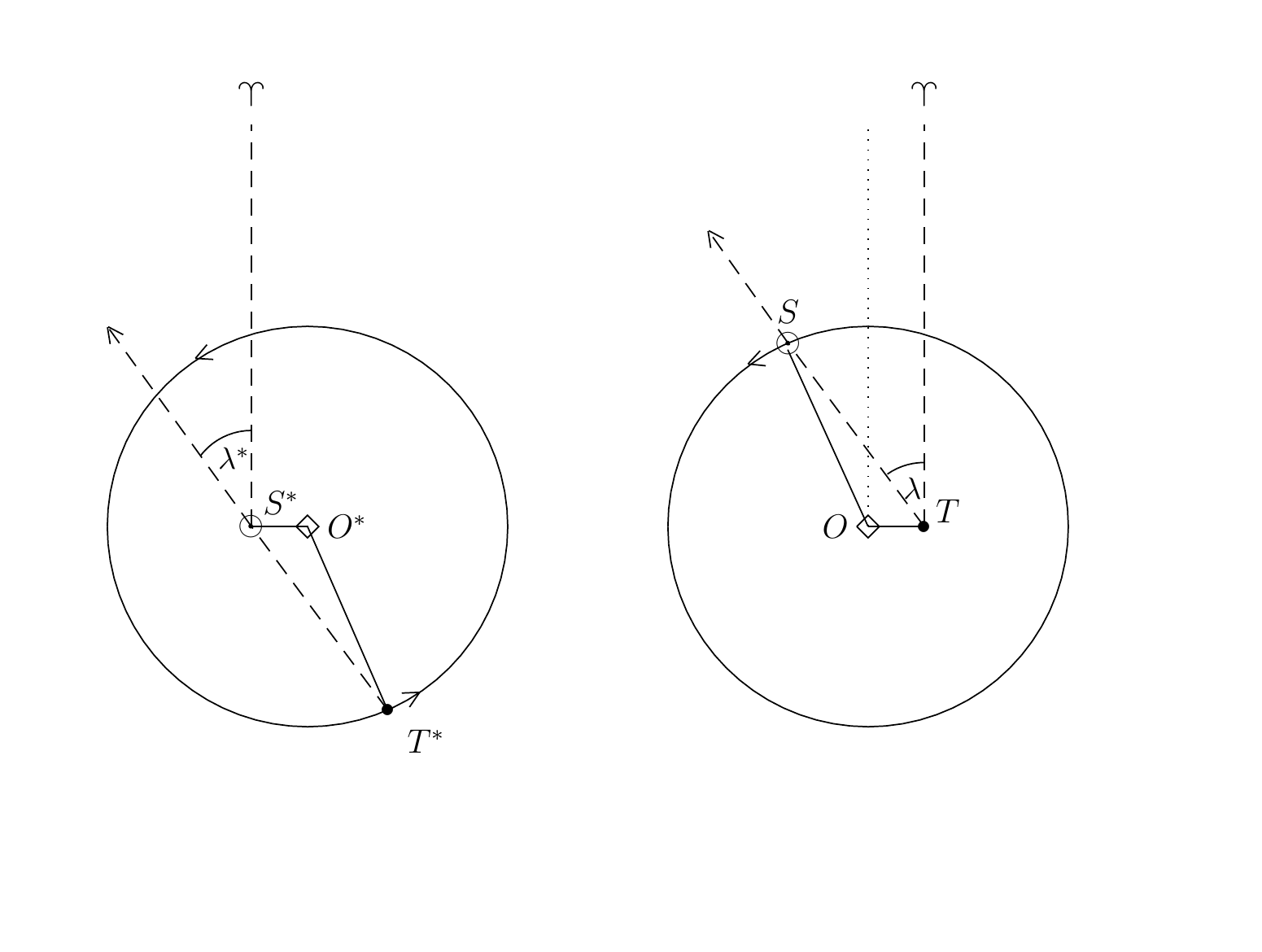}
\caption{Sun anomaly represented by decentered sun. 
Left: Earth $T^{\ast}$ moving uniformly on a circle  about the center $O^{\ast}$ with an eccentric sun $S^{\ast}$. Right: Kinematic inversion with exchange of the role of $T$ and $S$. Here  $\Delta OTS \sim \Delta O^{\ast} S^{\ast}T^{\ast}$ with  corresponding sides parallel: $OT \parallel   O^{\ast}S^{\ast} \, , \;  TS \parallel S^{\ast}  T^{\ast}\, \;   SO \parallel T^{\ast} O^{\ast} $. $S$  moves  uniformly about $O$, while for an observer in $T$ its motion is non-uniform, i.e, the sun's ecliptical longitude $\lambda(t)$ no longer progresses proportional to the time. \label{fig ecc S-T}}
\end{figure}

The  {\em simplest case} and first step to do  is, of course, to  translate the    motion of the earth $\earth$  ({\em terra}) about the sun  $\astrosun$
 ({\em solis}). To do so one needs a geometric imagination of  the ``true  motion''\footnote{Here and in the following the attribute {\em true motion} is used in an Aristarchian heliocentric perspective; in the following it will no longer be relativized by  using of  quotation marks.}
 which can be represented in a Euclidean plane by a similar picture in the strict geomerical sense (see figure  \ref{fig S-T}, left, for the motion of the earth about the sun).\footnote{A  more detailed picture can be found in  \citep[fig. 8]{Aaboe:Episodes2}.}  Here and in the following  symbols labelled by an asterisk like $S^{\ast}$, $T^{\ast}$ denote (moving) points in the plane representing the true objects, here the sun and the earth. The corresponding symbols without asterisk in the figure  on the right represent a geometrical image of the apparent motion  as  seen from the earth $T$. Here, of course,    $S$  is the moving point. Because the  practically ``infinite'' distance to the  vernal point $\aries$ ({\em aries}), the ecliptic locus marking the position of the sun at the spring equinox, the  broken lines (indicating the direction from $S^{\ast}$  to $\aries$, respectively from  $T$ to $\aries$) are parallel; for the connecting lines $T^{\ast}S^{\ast}$ and $TS$ the same holds, including orientation if one wants. Thus the respective angles are equal,  $\lambda^{\ast}(t)=\lambda(t)$ (a simple exercise in similarity geometry). The translation results in a  motion  of the sun $S$ along the ecliptic, which is {\em kinematically similar} to the motion of the earth $T^{\ast}$ about $S^{\ast}$ in the following sense: 
\begin{itemize}
 \item[--] Angles at a given time are equal, $\lambda(t)=\lambda^{\ast}(t)$, and  proportions are conserved, i.e., the ratios of distances in the figure on the right (the picture for a terrestrial observer) are equal to those of the corresponding distances on the left (similar to the true motion). 
	 \item[--] In particular uniformity  of circular motions  and  their periods are conserved. For the case of disuniformity similarity still holds in the sense of $\lambda(t)=\lambda^{\ast}(t)$ (see fig. \ref{fig ecc S-T}). 
	 \item[--] The role of $T^{\ast}$ and $S^{\ast}$ is being  {\em inverted} in the pair $S$ and $T$.
	\end{itemize}
For those   accepting a heliocentric hypothesis (perhaps even sharing the heliocentric ontological conviction)  such a geometrical consideration {\em explains} the apparent motion of the sun in the course of the year through  the ecliptic circle, independent of the question of  quantitative precision. Moreover this simple example makes clear in which sense we encounter here a  ``kinematic inversion'' from true to apparent motions.

 But a  warning is appropriate: A  kinematically inverted picture of the earth's uniform circular motion about the sun does not do justice to  the unevenness of the sun's  motion in the course of the year (the ``sun anomaly'').  The sun anomaly 
 had long been known in Greek science, probably already in the Miletian period. It was  well documented at the time of Eudoxos  \citep[p. 295]{Szabo:Astronomie} and   accounted for  in the homocentric sphere model by Kallippos about the  middle of 
the 4th century BC by an ad hoc  introduction of additional spheres.\footnote{Kallippos is reported to have used the data $(95,92,89,90)$ for the length of the seasons, counted from the vernal equinox  \citep[p. 106]{Dreyer:Astronomy}}
In the above form the  heliocentric picture could not yet compete with the refined  homocentric sphere approach of Eudoxos and Kallipos. 
But also here an   ad hoc   method, even a simpler one than in the homocentric model,  may be used to  mend the deficiency. It  
 consists of a displacement of the true sun from the center of the earth's motion which is still  assumed as uniform and circular with regard to a center $O^{\ast}\neq S^{\ast}$.
A similar  decentration
 is then obtained for the apparent motion of the observable sun $S$ which moves  uniformly with respect to a center $O$ such that $\Delta TSO \sim \Delta S^{\ast}T^{\ast}O^{\ast}$ (orientation of edges changed) (figure \ref{fig ecc S-T}). This results in  an eccentric circular motion of which we know that it was studied by Apollonios.   For the planetary motion this type of correction implies a simple {\em  eccentric deferent model} (figure \ref{fig exc deferent}).  


The uniform motion of $S$ about $T$ in fig. \ref{fig S-T} reflects  the motion of the so-called  {\em mean sun}.  Although Greek astronomer knew well  that uniform motion is not what one sees at the sky, one can still work with it as 
 an idealization arising from averaging. The observational position of the sun  deviates from the mean sun; it was  called the {\em true sun} (in a second, phenomenological meaning). For disambiguating symbols, one may prefer to introduce different letters for the mean sun $S_m$ and the observationally ``true'' sun which we will denote by $S_{\astrosun}$,  where necessary. 

\subsection{\small Side-remark on the order of the planets \label{subsection order of planets}}
In the homocentric system of spheres the order of the arrangement of the planets was a conventional  question and changed with different authors. A heliocentric view, on the other hand, will immediately see a qualitative difference of the kinematic appearance between planets with orbital radius smaller than that of the earth and those of  larger radius. Those of the first group stay  kinematically close to the sun, seen from the earth, i.e. they never can form an angle $\alpha > 90^{\circ}$ (a right angle $R$ in early Greek terminology), the latter will take on any angular distance to the sun, with the extreme positions of opposition ($\alpha= 180^{\circ}=2R$) and conjunction ($\alpha=0$) to the sun. The planets of the first group ( Mercury $\mercury$ and Venus $\venus$) are called the ``inferior planets'' by historians of astronomy who try to avoid the heliocentric connotation of ``inner''. The second group (Mars $\mars$, Jupiter $\jupiter$, Saturn $\saturn$) are then called the ``superior planets'',  avoiding the qualification as ``outer'' ones.\footnote{The terminology ``inner'' and ``outer'' planets is used  in \citep{Neugebauer:HAMA}}
 The anomalies of the motion on the ecliptic (in particular the maximal elongations of the inferior planets from the sun, and for the superior planets from its mean position) suffices for establishing an order of increasing orbital radii, already on the basis of rough data on the change of planetary positions over the course of the year. 
Although the terminology  {\em inferior, superior} is not to be found in Ptolemy's {\em Almagest}, he made a clear distinction between the two groups and proposed a definite order of the  ``spheres'' of the planets and the sun indicated by (Book IX, 1):
\beq \mercury \quad \venus \qquad \astrosun \qquad \mars \quad \jupiter \quad \saturn \;  \label{eq order of planets}
\eeq
This is the  heliocentrically ``correct'' order if one keeps the kinematic inversion of the role of sun and earth in mind. Ptolemy {\em gave no systematic reason} for this order, although he could have given one by the order of the respective radii of deferents. He rather preferred  to  
give  a slightly confused  {\em  historical justification} for the placing of Venus and Mercury below the sun.\footnote{``But concerning the spheres of Venus and Mercury, we see that they are placed below the sun's by the more ancient astronomers, but by some of their successors these too are placed above [the sun's], for the reason that the sun has never been obscured by them [Venus and Mercury] either.'' \citep[p. 419]{Ptolemaios/Toomer:Almagest}} 
All this looks very much like a hidden (forgotten or suppressed) influence of an input of   heliocentrism into the tradition of planetary theory in Hellenistic times.\footnote{This impression is strengthened, against the intention of the author, by O. Neugebauer's discussion of the question in his {\em History of Mathematical Astronomy}. He over and over states that heliocentric investigations could not have been of much influence on the development of Greek planetary theory because, according to him, no quantitative conclusion could be drawn from them (see sec. \ref{section discussion}). 
But the information he gives on the sources seem to speak against this picture.
In an extended presentation of source material diverse different arrangements of the planets in Greek and Mesopotamian astronomy \citep[IV C 2.1, \S 2.1]{Neugebauer:HAMA} 
he comments  Ptolemy's order (\ref{eq order of planets}). He remarks that Cicero (106--45 BC), Vitruvius (1st c. BC),  Pliny (23--79 AD), and Plutarch (45--125 AD) ``took it for granted'' \citep[p. 691]{Neugebauer:HAMA}. With other words,   the order of planets stabilized to the one indicated by a heliocentric view in the time after Hipparchos. This is no definitive proof of a heliocentric influence on Greek planetary theory but adds to the circumstantial evidence for it. 
}

\subsection{\small Kinematic inversion of the motion of inferior planets \label{subsection inferior planets}}
\begin{figure}[h]	
\hspace*{2em}	\includegraphics[angle=0,scale=0.7]{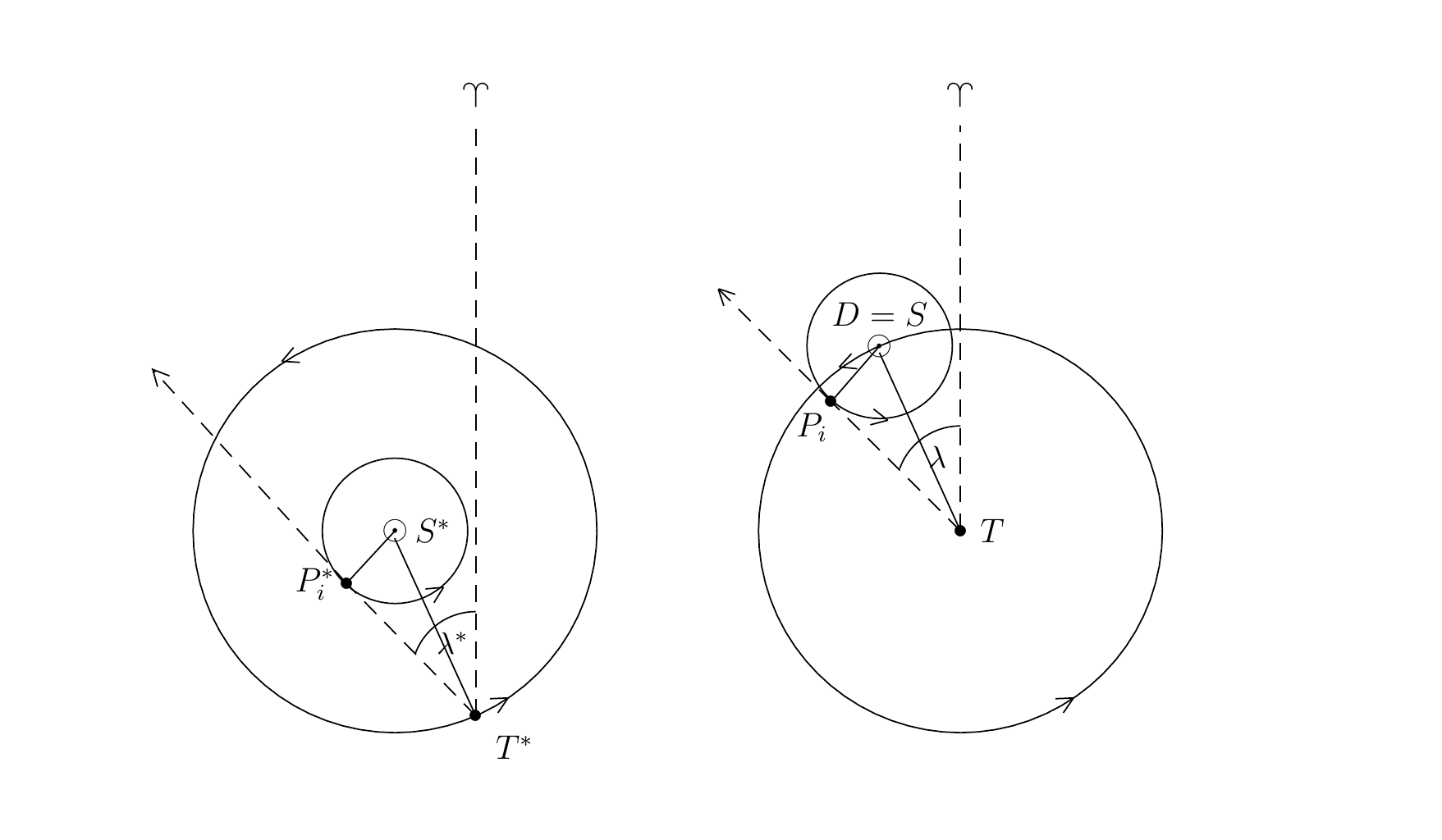} 
\caption{Kinematic inversion for inferior planets. Left: 
Motion of an inferior planet $P_i^{\ast}$ and the earth $T^{\ast}$ about the sun $S^{\ast}$.  Right: Apparent motion of the planet $P_i$ seem from a terrestrial observe at $T$. $P_i$ moves   on an epicycle with center $D$ identical to the position of the apparent sun ($D=S$). The similarity of triangles $\Delta T^{\ast}P_i^{\ast}S^{\ast} \sim \Delta TP_iS$  is due to the  parallelity of  corresponding sides. $\lambda(t)$ is the time dependent ecliptic longitude of the planet. \label{fig S-P-T}}
\end{figure}

The kinematic inversion of the motion of planets $P^{\ast}$ with true orbit inside the   path of the  earth $T^{\ast}$ (Mercury $\mercury$ and Venus $\venus$) is nearly as simple as that for the earth itself, if one assumes concentric circular true orbits of $P^{\ast}$ and $T^{\ast}$.
The apparent  motion of such a planet is given by a point $P$ moving on a  circle about a center, in the terminology of epicyclic models called the {\em deferent}  $D$ (figure \ref{fig S-P-T}). The geometrical construction resides on parallel and  similar triangles with  directed sides. The deferent  moves with the apparent position $S$ of the sun, constructed in the first step (figure \ref{fig S-T}). Because   corresponding connecting lines are parallel, $TD \parallel T^{\ast}S^{\ast}$, $DP \parallel S^{\ast} P^{\ast}$ and $TP \parallel T^{\ast} S^{\ast}$ the kinematic picture (fig. \ref{fig S-P-T} right) is again similar to the true geometric constellation (fig. \ref{fig S-P-T} left). 
In particular the ratios formed by the radii $r_P=|S^{\ast} P_i^{\ast}|,\, r_T=|S^{\ast} T^{\ast}|$ and the radii
 $r_1=|TD|, \, r_2=|D P_i|$ of the inverted picture are equal:
 \[ r_P:r_T = r_2:r_1
 \]

 The description of the apparent motion of interior planets (under the  assumption of concentric true orbits) {\em leads necessarily to a simple epicyclic presentation}.    For heliocentrists such epicycles are, of course,   not ``real``  but only constructions devised to  ``save'' (express) the phenomena, i.e., they are  phenomenological models in the modern terminology.
  
\subsection{\small Superior  planets, qualitative kinematics \label{subsection superior planets qualitative}}
\begin{figure}	
\hspace*{5em}	\includegraphics[angle=0,scale=0.7]{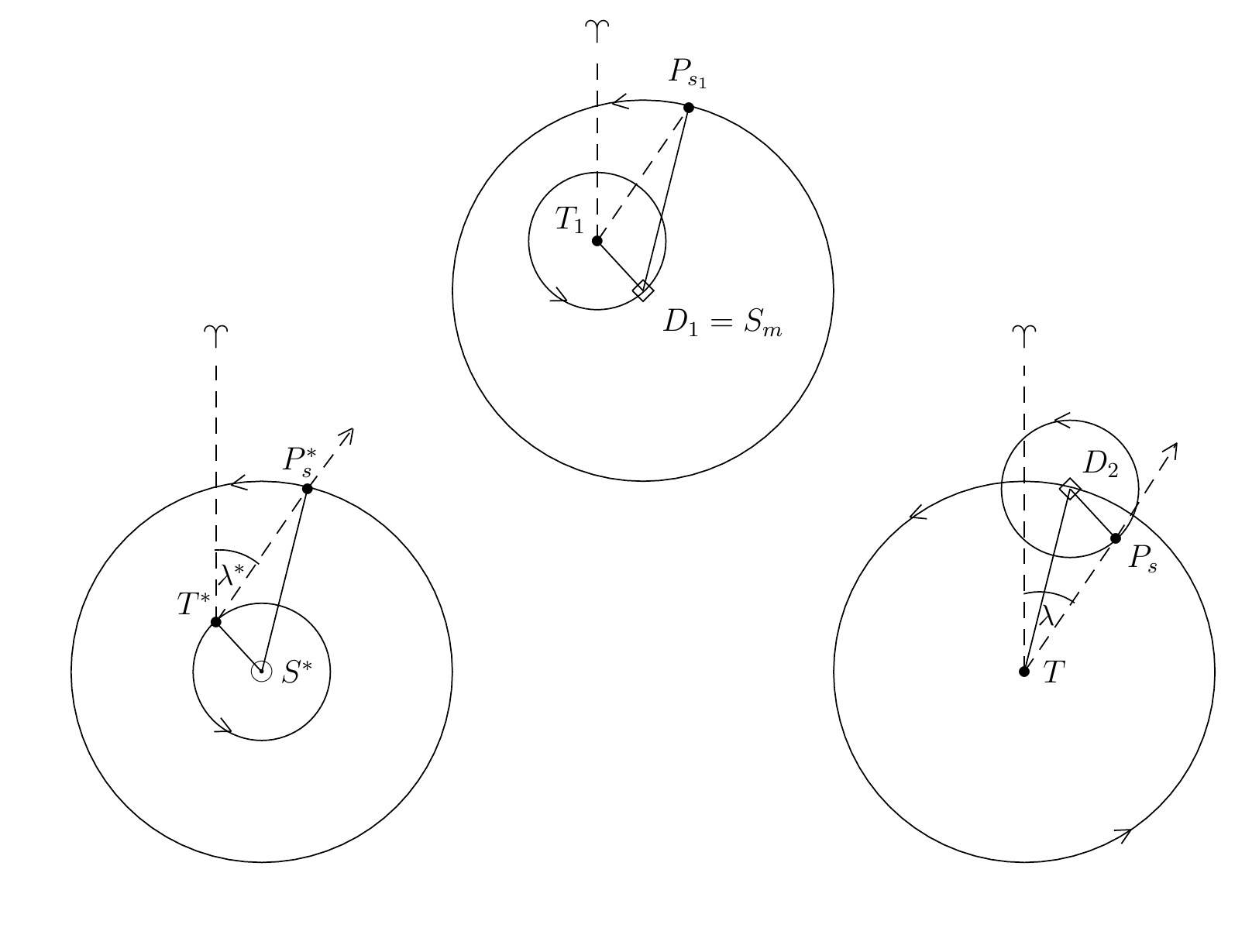} 
\caption{Kinematic inversion of the motion of the earth $T^{\ast}$  and a superior  planet $P_s^{\ast}$ about the sun $S^{\ast}$. Left: True motion  about $S^{\ast}$. Middle: First kinematic inversion like for inferior planets with deferent identical to the mean sun $D_1=S_m$ (cf. fig. \ref{fig S-P-T}). This leads to a larger radius of the epicycle than for the  deferent. 
Right: Second kinematic inversion with interchanged radii. The  deferent $D_2$ moves now on the larger circle, the planet $P_s$  as seen from an observer on $T$  on an  epicycle of smaller radius. Note the similarity of triangles, in particular $\Delta T^{\ast}S^{\ast}P_s^{\ast} \sim \Delta P_sD_2T$ (see text). Here motion of the deferent $D_2$ reflects  the  motion of $P^{\ast}$ about $S^{\ast}$, while the epicyclic motion of $P_s$ about $D_2$ reflects the  motion of the earth $T^{\ast}$  about the sun $S^{\ast}$. (Ecliptic longitude $\lambda(t)$ in this picture is being charted to the West.) \label{fig S-T-P}}
\end{figure}

After the first kinematic inversion of the motion of the sun as above, a superior planet $P_s$ will   run on an epicycle with larger radius than that of the deferent ($D_1=S$) (figure \ref{fig S-T-P}, middle). 
Mathematically this is no problem. Basically this is all one needs for a geometrical representation of the superior planet's motion   seen from the earth. 
For a graphical (or later quantitative) evaluation  it may be a   disadvantage that the deferent $D_1=S$ has a smaller period (and thus greater angular velocity) than the motion of $P_s$ on its epicycle. The constellation can just as well be understood as  a circular motion of $P_s$ about a variable eccenter $D_1$ moving on a circle with center  $T$. 

This seems puzzling and  one may look for an equivalent  kinematic representation which  is closer in structure to the one for the inferior planets. A similarity consideration of moving triangles allows to transform the kinematics of $P_{s_1}$ about $T$ (fig. \ref{fig S-T-P}, middle) into one given by an epicyclic motion with  deferent radius $|T D_2|$  larger than the radius $|D_2P_s|$ of the epicycle. Figure \ref{fig S-T-P}, right, shows how this can be achieved by  a  second kinematic inversion  by which  the roles of deferent and epicycle are exchanged.

  In this figure the following parallelisms are demanded to hold at each moment of the  motion:
  \[ D_1P_{s_1} \parallel S^{\ast}P_s^{\ast} \parallel T D_2, \quad D_1T_1 \parallel S^{\ast} T^{\ast} \parallel P_s D_2 , \quad   T^{\ast}P_s^{\ast}  \parallel  TP_s \;\;    \;
 \]
Then the following moving triangles are similar:\footnote{In vector notation with similarity factor $\lambda$, Def/construction: $\stackrel{\longrightarrow}{TD}=\lambda \stackrel{\longrightarrow}{S^{\ast}P^{\ast}}, \quad \stackrel{\longrightarrow}{DP}= \lambda\stackrel{\longrightarrow}{T^{\ast}S^{\ast}}$; 
thus $\stackrel{\longrightarrow}{TP}\,=\,\stackrel{\longrightarrow}{TD}+ \stackrel{\longrightarrow}{DP} = \lambda(\,\stackrel{\longrightarrow}{S^{\ast}P^{\ast}} + \stackrel{\longrightarrow}{T^{\ast}S^{\ast}})\,=\, \lambda \stackrel{\longrightarrow}{T^{\ast}P^{\ast}}$. This can, of course, been  expressed  in terms of classical geometry like in the main text.}
 \[ \Delta T_1 D_1 P_{s_1} \; \mbox{(middle)} \;  \sim   \Delta T^{\ast} S^{\ast} P_s^{\ast} \; \mbox{(left)}   \sim \Delta P_s D_2 T \; \mbox{(right)}
 \] 
If one keeps the similarity ratios constant during the motion, $D_2$ lies on a circle about $T$ and $P_s$ on a (moving) circle about $D_2$. 
  Note also the direction of edges.  Expressed in modern terminology, the orientation of the  triangle on the right has been changed with regard to the  other ones (left and middle).
  
 Kinematically the configuration of fig.~\ref{fig S-T-P}, right,  implies an apparent  motion of $P_s$, satisfying the following conditions:
 \begin{itemize}
 \item[(i)] $D_2$ moves on a circle about $T$ similar to the motion of $P_s^{\ast}$
 about $S^{\ast}$ (with the same orientation and angular velocity).
 \item[(ii)] $P_s$ moves on a circle about $D_2$ similar to the motion of $T^{\ast}$ about $S^{\ast}$.
 \item[(iii)] The planet is in opposition to the sun (i.e. $P_s^{\ast},\, T^{\ast}, S^{\ast}$ collinear) for $T, P_s, D_2$ collinear; i.e., in the ``middle'' of the retrograde motion of the planet on the ``lower'' arc of the epicycle. 
 \end{itemize}
Using the language of epicycles, the (superior) planet $P_s$ moves on an epicycle about the deferent $D_2$; but after the second kinematic inversion the epicyclic model gives a rather ``surreal'' picture of the real geometric constellations (e.g. by item (iii)). A typical  retrogradation loop in the apparent motion of  superior planet $P_s$ is generated by passing through the epicycle  once; in this time its mean position marked by $D_2$ has progressed along the ecliptic; this motion is called a {\em synodic cycle} of the planet. In the course of such a cycle specific {\em synodic events} play a distinguished role for observation and theory:  the opposition or conjunction to the sun, a (forward, respectively backward) stationary point in the loop of retrogradation, or the day of first or last visibility.

The directed  line 
{ $D_2P_s$} (deferent --- planet) is parallel (including direction) to $T^{\ast}S^{\ast}$ and thus to $TD_1$,  the  direction of the mean sun  (because of $D_1=S$). Condition (i) tells us then that the epicyclic motion of the planet  mirrors the  motion of the earth about the sun respectively,  after the first kinematic inversion, the motion of the  mean sun. 
This property was still demanded by the construction in Ptolemy's {\em Almagest};
 but there it appeared as an {\em  unexplained and surprising feature} inbuilt into the formal rules of  epicycle construction (cf. section \ref{section discussion}).  
Similarly condition (ii) means that  the motion of the deferent mirrors the averaged (mean) motion of the planet about the sun. {\em This feature could not even be expressed} in the framework of Ptolemaos' { reduced epicylic framework}, although $TD_2$ was miraculously interpreted as as a representative for some kind of mean distance to the planet. His epicyclic method was 
{\em  reduced} in the sense that the heliocentric origin of the construction was no longer part of  contemporary  knowledge;  at least as  regards the  parts  which were transmitted to later centuries  (see section \ref{section hist hypothesis}).

\subsection{\small First quantitative adaptation to observational data and a provisional resum\'e \label{subsection superior planets quantitative}}
 While the historical knowledge on Mesopotamian astronomical data and algorithms can now rely on detailed source material \citep{Neugebauer:HAMA}, this is not the case for classical and Hellenistic Greek astronomy. This  may have  contributed to  the general view  that numerical evaluation of geometrical models using  data  of Mesopotamian astronomy did not start effectively before  Hipparchos, who was also  the central figure for introducing sexagesimal numerics and trigonometric tables  into Greek applied mathematics  \citep{Jones:1990,Jones:1991}. We know, however,   that theoretical questions which are important for the  quantitative adaptation of  epicyclic models to observational data were already topical   at the time of Apollonios. 
It is thus not convincing to assume that astronomer-mathematicians of the late 3rd c. BC have {\em not} undertaken attempts for such quantitative adaptations of the models. 

Regarding  observational data on planetary motion   in pre-Hipparchian times, Greek astronomers could start from a  rudimentary knowledge  of the number count of  synodic cycles $s$  and the associated number $r$ of revolutions  of a  planet in the ecliptic during a given number $N$ of solar years. They were  recorded by Mesopotamian and/or Egyptian scribes for {\em long periods} of the planets.\footnote{A ``long period'' is the time between two events in which the planet stands at the same place in the ecliptic in  equal synodic constellations. A ``synodic constellation'' may be specified by any of the distinguished synodic events.}
This suffices for a first quantitative adaptation of the epicyclic models to the observations by considerations  which were later refined by Hipparchos. Later they found entrance into the corpus of the {\em Almagest}. The procedure is  explained in terms of Mesopotamian sexagesimal calculations in \citep[p. 79ff.]{Aaboe:Episodes2}. The same method can just as well be expressed in terms of the classical  Greek concept of angle quantities and the calculus of numerical ratios, i.e., with the methods available already  during the 3rd c. BC.
 
\begin{figure}[h]	
\hspace*{8em}	\includegraphics[angle=0,scale=0.8]{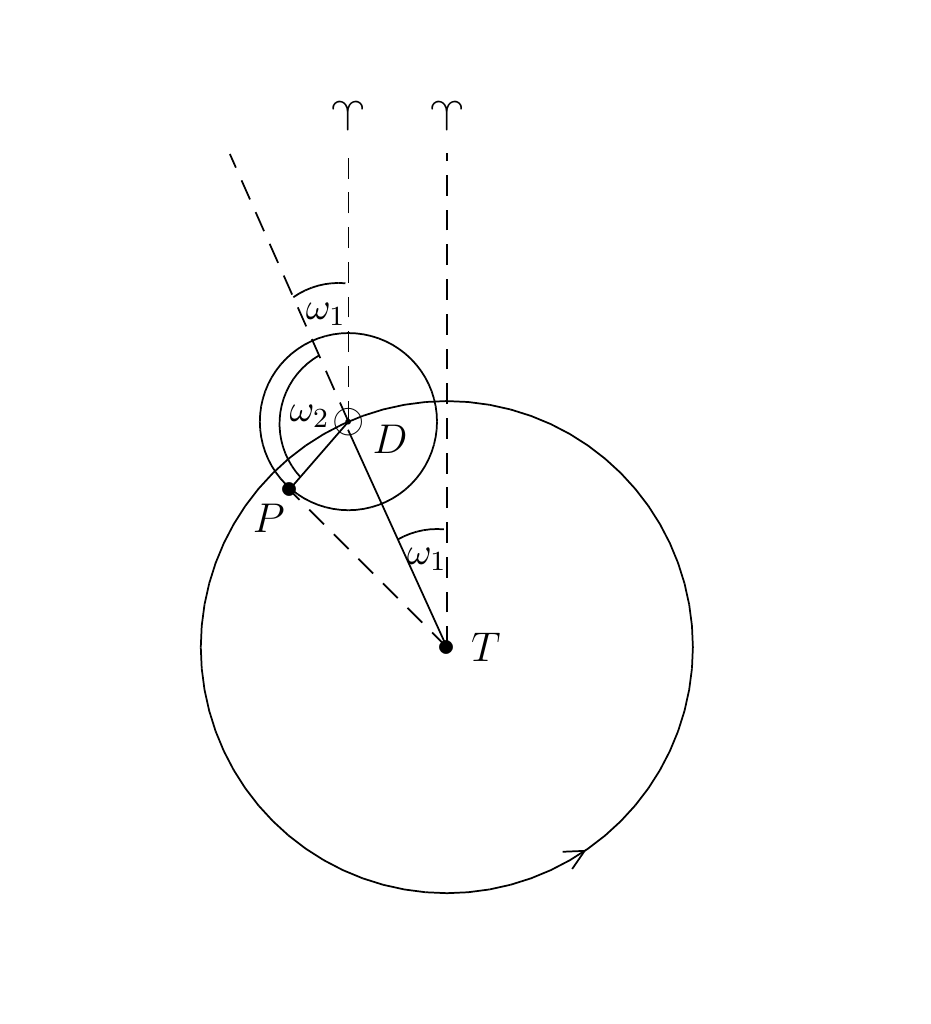} 
\caption{Progression of angles in epicyclic model: The deferent $D$ progresses by angle $\omega_1$ with regard  to a fixed reference direction on the ecliptic (here aries $\aries$).
The  planet $P$  moves on an epicycle with angle  $\omega_2$ in relation  to the radial direction of the deferent. The angle $\omega$ of the planet $P$ with regard to $\aries$   (and its progression) is $\omega = \omega_1 + \omega_2$. \label{fig angles}}
\end{figure}

 In the 4th c. B.C.  the arc of the regular dodecagon, also called a   {\em zodion} $z$ (lat. {\em signum}, i.e., sign of the zodiac) was sometimes used as a basic angle unit   ($1 z = 30^{\circ}$).\footnote{The zodion was used by Autolykos (4th c. B.C.)  \citep[pp. 92f.]{Autolykos}, cf. \citep[p. 99]{Szabo:Entfaltung}, also \citep[p. 593]{Neugebauer:HAMA}.} 
With  a set of data $(r,s,N)$ as above for a superior planet $P_s$, $r$ number of ecliptic revolutions, $s$ of synodic cycles in $N$   years,  the mean angular progression $\omega_1$ of $P$ per day ($d$) in the ecliptic (in later terminology the angular velocity in units $z\, d^{-1}$) is given by the ratio
\[ \omega_1 = \frac{12\, r}{N\cdot Y}  \quad \mbox{with} \; Y= \mbox{number of days per year (usually}\; 365\frac{1}{4})\, .
\]
Similarly for the progression $\omega_2$ (angular velocity in units $z\, d^{-1}$) of the  deferent $D_2$,  the ``mean motion'' of the planet is 
\[ \omega_2 = \frac{12\,s}{N \cdot Y}\, .
\] 
As the progression of $P_s$ on the epicycle is measured against a synodic reference line, e.g. the one defining the conjunction of the planet with the mean sun, the progression $\omega$ of $P_s$  
with regard to a fixed reference point of the ecliptice, e.g. aries $\aries$, is given by
\[ \omega= \omega_1 + \omega_2 \qquad \mbox{(fig. \ref{fig angles}).}
\]

 For an outer planet $\omega$ is the progression of the mean sun, which is 1 round angle per year, or $\frac{12}{Y}$ in terms of zodiacs per day. The heliocentric hypothesis leads thus to the constraint $\frac{12}{Y}=\frac{12(r+s)}{N \cdot Y}$, i.e.,
 \beq r + s = N \quad \mbox{for superior planets.}  \label{eq superior planets}
 \eeq
Moreover, it is an immediate consequence of the heliocentric assumption  that  
\beq r=N \quad \mbox{for inferior planets.} \label{eq inferior planets}
\eeq
 These are   {\em structural predictions of the heliocentric hypothesis}, which  could easily be stated with the knowledge of Greek astronomy of the 4th century BC.  Due to the lack of sources we cannot say which Mesopotamian data were available to Greek astronomers at this time, but it  is clear that  (\ref{eq superior planets}), (\ref{eq inferior planets}) are  consistent with  Mesopotamian observations and also with the later  Greek ones. 
Such data were contained  in Babylonian (``goal year) texts of the 4th c. BC.\footnote{\citep[p. 151]{Neugebauer:HAMA},\citep{Gray/Steele:2008}.} 
Much later they were reproduced in the {\em Almagest}, vol. IX.3,  by Ptolemy  who ascribed them to Hipparchos. It seems likely that  such simple data sets were already known to Greek astronomers of the 3rd c. BC. In the following table Ptolemy's data $(N,r,s)$ are   given in simplified form,  rounded to integer numbers.   The corresponding values for $\omega_1, \omega_2$ in $z/d$ (zodion per day) have been added in the Greek style as continued fractions up to order 2 ($[a_0; a_1,a_2]= a_0+1/(a_1+1/a_2)$), in decimal fractions ($^{\circ}/d$) and in total revolutions per year.

\vspace{0.2cm}
\begin{center}
\begin{small}
\begin{tabular}{|l||c|c|c||c|c|c|c|}
\hline 
 & N & r & s & $\omega_1\; (z/d) $ & $\omega_2 \; (z/d)$ & $(\omega_1, \omega_2)$\; ($^{\circ}/d $)&  $(\omega_1, \omega_2)$\; rev.$/y$ \\
 \hline 
 $\mercury$ Mercury  & 46 & 46 &145  & [0; 30,2]& [0; 9, 1] & (0.986, 3.107) & (1, 3.15)\\
 $\venus$  Venus & 8 & 8 & 5 &  [0; 30, 2]  & [0; 48, 1] & (0.986, 0.616)& (1, 0.625)\\
 $\mars$  Mars & 79 & 42 & 37 & [0; 57, 3]& [0; 64, 1] & (0.524, 0.462)& (0.532, 0.468)\\
 $\jupiter$ Jupiter & 71 & 6 & 65 & [0; 360, 5]& [0; 33, 4]& (0.083, 0.902)& (0.085, 0.915)\\
 $\saturn$ Saturn & 59 & 2 & 57& [0, 897, 1]  & [0; 31, 1]& (0.033, 0.952) & (0.034, 0.966)\\
\hline
\end{tabular}
\end{small}
\end{center}
Obviously the identities (\ref{eq superior planets}), (\ref{eq inferior planets}) are satisfied.\footnote{For a discussion of the Mesopotamian background of these data see  \citep{Jones/Duke}. See also  \citep[p. 79f.]{Aaboe:Episodes2} for the corresponding values for Venus and Mars in sexagesimal notation.}

	 For Aristarchos, or any of his contemporaries and successors studying the  heliocentric hypothesis,   the question arose how  the motion of celestial bodies appears to a terrestrial observer. The answer for the mean motion of the sun and for the planets  must have led a geometer of the time, {\em who posed the question},  to an insight boiling down to the  kinematic inversion of secs. \ref{subsection earth motion}, \ref{subsection inferior planets}, \ref{subsection superior planets qualitative}.
	 	 On a  qualitative level the retrograde motions of planets are well  {explained} by kinematically  inverting a heliocentric picture of the motion of the earth  and the  planets,  already under the most simple (and in a first approach ontologically appealing) assumption of {\em uniform  motion} of the planets including the earth, on concentric circles about the sun. And even  quantitative predictions  (\ref{eq superior planets}) and (\ref{eq inferior planets})  could be derived from the heliocentric assumption  translated to the corresponding epicyclic kinematics. In Mesopotamian astronomy the corresponding relations were known to hold by empirical reasons;\footnote{\citep[p. 312]{Neugebauer:Essays} calls   relations (\ref{eq superior planets}), (\ref{eq inferior planets})  ``empirical fact(s)'' and at another place as ``well known in Babylonian astronomy''  \citep[p. 389]{Neugebauer:HAMA}.} 
	 	   in the later Ptolemaic codification they appeared as formal rules without reason.   
	 	   
	 An estimate of the  relative radii $r_1=|TD|,\,  r_2=|DP|$ of the representing circles from empirical data was an important step for the adaptation of heliocentric and epicyclic models to the data. It could be  achieved in specific cases by  chord-calculations without the recourse to  general chord tables on the observational basis of   maximal elongations. The latter are easier to observe   than   stationary points, at least for inferior planets, cf. {\em Almagest}, IX.2. 

\begin{figure}[h]	
\hspace*{8em}	\includegraphics[angle=0,scale=0.8]{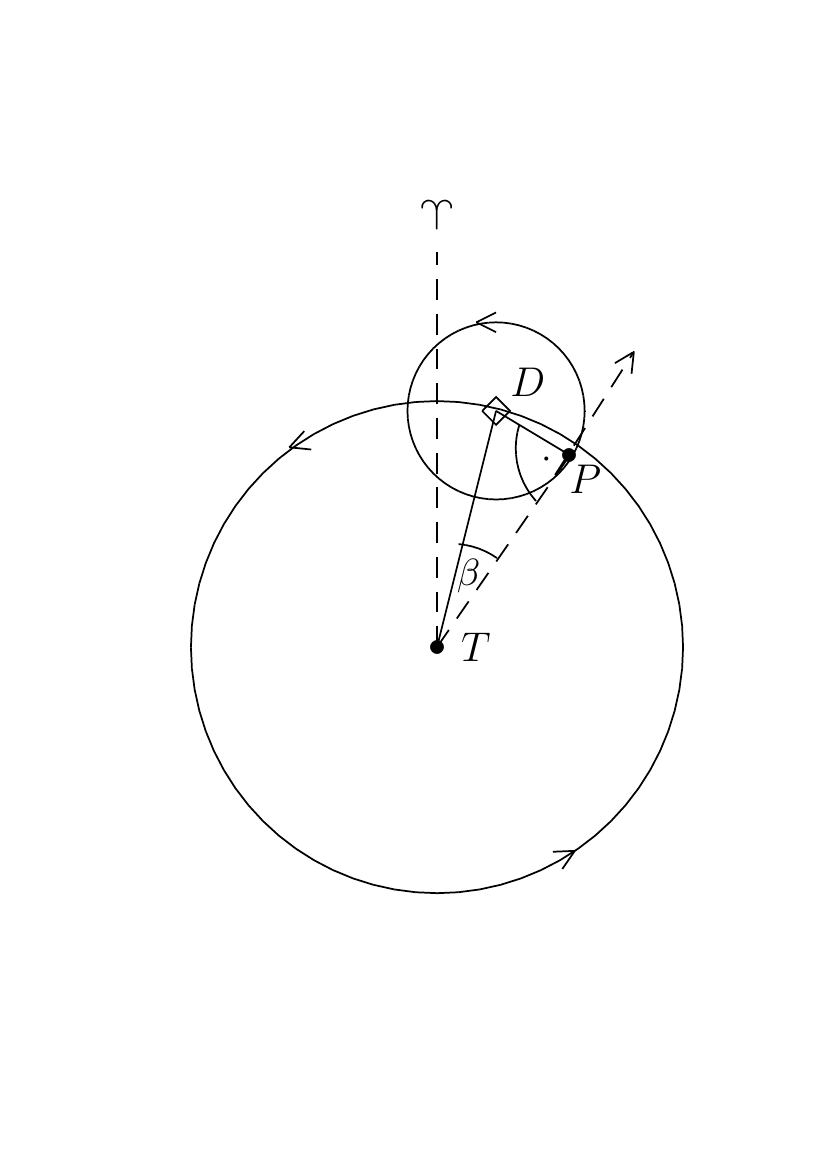} 
\caption{Maximal elongation $\beta$ of a planet $P$ on epicycle about $D$,  seen from earth $T$. \label{fig elongation}}
\end{figure}

If $\beta$  is the maximal elongation angle between a planet $P$ and the deferent $D$ (the mean sun for inferior planets and  the mean planet for the superior ones),  $r_1$
  the radius of the deferent circle and $r_2$ the radius of the epicycle radius, one finds (fig. \ref{fig elongation})\footnote{This  presupposes, of course,  centered deferent circles; in case of eccentricity modifications have to be made.}  
  \beq r_2= r_1 \sin \beta , \quad \mbox{or} \quad  2 r_2= r_1\,   ch\, 2 \beta \quad \mbox{in terms of chords $ch\, x$.} \label{eq radii}
  \eeq
 For the inferior planets (and if one is interested primarily in epicyclic modelling) one may like to  choose $r_1=1$ (radius of earth orbit as unit), or  $r_1= 60^p$ ($^p$ for partes) in Ptolemaic units. 
 For the heliocentric geometry of a superior planet one better sets $r_2=1$, respectively $60^p, $  (i.e., the kinematically inverted radius of the earth's orbit is taken as unit) and  to solve for $r_1$, the radius of the planet's orbit. 	   
  
  For interior planets the maximal elongation angle $\beta$ as evening or as morning stars has   been observed quite early  by Greek astronomers.  For Mercury  Ptolemy reports  data taken in 262 BC  ({\em Almagest}, IX.7) with  a mean of $\beta\approx 25^{\circ}$, while  
  two data taken by himself (or his group) ({\em Almagest}, IX.9) have a   mean $\beta\approx 23^{\circ}$. In pre-Hipparchian times, when chord tables were not yet established, one could   approximate this elongation by half the angle of a regular octogon ($\beta=22.5^{\circ}$). Then (\ref{eq radii}) leads to $r_2 : r_1 \approx 0.38 \sim 23^p$.\footnote{Ptolemy's own value for Mercury is $r_1=22^p30'$ ({\em Almagest} IX.10).} 
 Ptolemy reports elongation data for Venus taken by Timocharis in 272 BC ({\em Almagest}, X.4).\footnote{\citet[p. 130]{Waerden:1984} conjectures that the data taken by Timocharis, like those of other astronomers discussed in his paper,  were {\em intended} to determine the parameters of  a heliocentric planetary model.}
  He translated them into the angle to the mean sun,  with mean $\beta \approx 42.5^{\circ}$. In  pre-Hipparchian considerations this could be described  roughly as half the angle of the regular quadrangle ($\beta \approx 45 ^{\circ}$), which implies $r_2 : r_1 \approx 0.71 \sim 42^p$.\footnote{In his refined model Ptolemy arrived at $r_2=43^p 10'$.}

Observations of the maximal elongation of superior planets from their mean position play no role for Ptolemy's approach; this may be the reason that they are  not recorded by him. But once one has the  picture of the second kinematic inversion of superior planets in mind (fig. \ref{fig S-T-P}, right), it is clear how to  observe it: One may, e.g., start from the opposition of the planet (the position of  $P_s$  collinear with $T$ and $D$ on the lower arc of the epicycle) at ecliptic longitude $\lambda_0$. Then one observes the retrograde motion of $P_s$  and continues to follow the path of the planet until, after the time $\Delta t$ (in days), it reaches the longitude $\lambda_1$, where its  motion is now prograde and in tune with the progression of the mean planet ($D_2$ in our notation). Then the maximal elongation $\beta$ has been reached. With $\Delta \lambda=\lambda_0 - \lambda_1$ and  the progression of $D_2$ given by $\omega_1$ and $\Delta t$ it is 
\beq \beta = \Delta \lambda +  \omega_1\, \Delta t\, . \label{elongation outer}
\eeq 
One expects that for large outer radii (Jupiter and Saturn) $\Delta t$ is roughly a quarter of a year; while for a planet with orbit not much outside the earth's orbit  (i.e., Mars)  $D_2$ has considerably progressed during this time,  and  $\Delta t$ may be considerably  longer than a  quarter year. Lacking historical records on such data, the application of this method is here being checked  on  ``simulated''  data.
With just one data set of 2022 for each outer planet (thus less balanced than taking mean values of $\beta$ starting from different positions of the opposition on the ecliptic)  we find (rounded to half degrees):\footnote{Opposition of Mars Dec 08, 2022, of Jupiter Sep 26, 2022, and of Saturn Aug 14, 2022. Ecliptical longitudes at opposition: Mars $52.03^{\circ}$, Jupiter $339.64^{\circ}$,  Saturn $297.79^{\circ}$; $\Delta \lambda$ for Mars $-26.25^{\circ}$, for Jupiter $4.2^{\circ}$, for Saturn $3.04^{\circ}$ after $\Delta t=$ 130 days for Mars, 90 days for Jupiter and  Saturn. For Saturn the angular velocity is so small that a naked eye observation of the exact day of maximal elongation  is impossible. But   the progression of longitude between  $\Delta t = 90$ and 100 is just $+0.23^{\circ}$ and $10 \omega_1 = 0.33$,  which leads to a negligible difference in (\ref{elongation outer}). Ephemeris data 2022 from internet resource \url{https://www.mpanchang.com/planets/ephemeris/}. This  (Vedic) astrological website uses  modern ephemeris calculations; it is useful  for our purpose because it displays ecliptical longitudes and ecliptical angular velocities of all planets  for arbitrary times. Those who prefer could even ``simulate'' the elongation observations for any date in the 3rd century BC.
}  
 \\
\hspace*{4em} for Mars $\beta \approx  42^{\circ}$, \quad for Jupiter $\beta \approx 11^{\circ} $ \quad  for Saturn $\beta \approx 6^{\circ}$\\
In pre-Hipparchian times one would  approximate chords of twice  these angles by regular $n$-polgyons, with $n=4$ for Mars ($\beta \approx 45^{\circ}$ like for Venus, but with reciprocal proportion), $n=16$ for Jupiter ($\beta \approx 11.25^{\circ}$), and $n=30$ for Saturn ($\beta \approx 6^{\circ}$). The corresponding values for the radii of the planets  expressed in terms of  the radius of the earth orbit are given in the following table; added are the proportions $r_2:r_1$ in {\em partes} and the corresponding Ptolemaic values; in the last column  the modern mean distances of the planets to the sun are given in  astronomical units $AU$.

\vspace{0.2cm}
\begin{center}
\begin{small}
\begin{tabular}{|l||c|c||c|c|c|}
\hline 
 & $r_1$ & $r_2 $ &  $r_2:r_1$ in $^p$ &  Ptolemy in $^p$  & distance in $AU$ \\
 \hline 
 $\mercury$ Mercury  & 1 & 0.38 & 23 &  22;30&  0.387  \\
 $\venus$  Venus & 1 & 0.71 & 42 & 43;10   & 0.723 \\
 $\mars$  Mars & 1.41 & 1 & 42 & 39;30 &1.523\\
 $\jupiter$ Jupiter & 5.13 & 1 & 12 & 11;30 & 5.202  \\
 $\saturn$ Saturn & 9.6 & 1 & 6;27 & 6;30 & 9.538 \\
\hline
\end{tabular}\\[1em]
\end{small}
\end{center}

For the  simplest possible approach to heliocentrically founded epicyclic astronomy   the quantitative results are surprisingly close to the values of Ptolemy's much more refined model and the modern values of the mean radii of the planetary orbits. It should be added that all the regular $n$-gons appearing here ($n= 4, 8, 16, 30$)  are  constructible in the Euclidean style, i.e., with ruler and compass.

\begin{figure}[h]
\hspace*{10em}\includegraphics[scale=0.7]{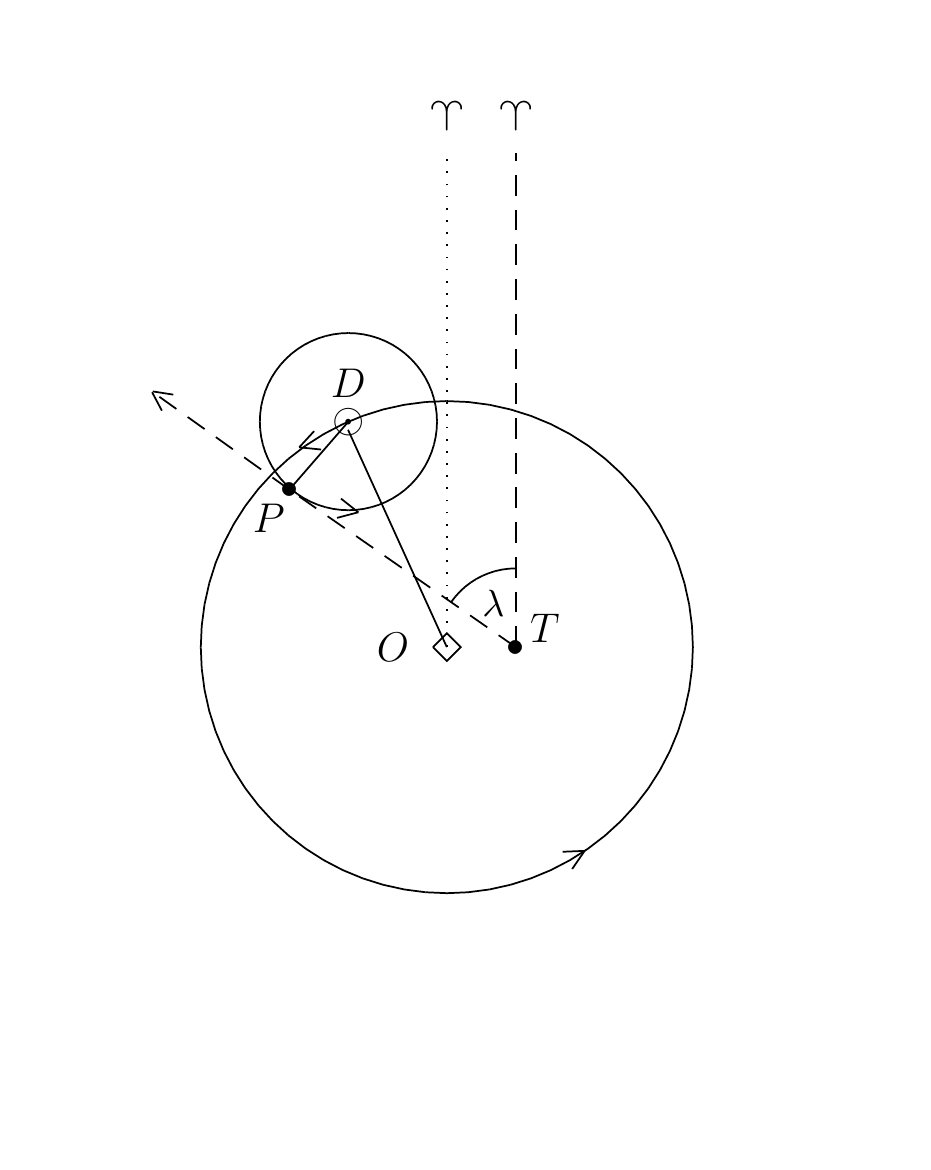}
\caption{Simply eccentric deferent model arising from kinematic inversion of the orbit of an inferior planet moving about a decentered true sun (cf. fig. \ref{fig ecc S-T}). The deferent  $D$ moves uniformly on a circle about $O$, but is seen with non-uniformly changing ecliptic longitude $\lambda$ for a terrestrial observer at $T$ (``ecliptic anomaly'' for  the motion of the mean planet).  \label{fig exc deferent} }
\end{figure}

But it must have been clear that planets do not move as uniformly as such an averaged picture  suggests. If one goes a bit deeper into quantitative detail, two types of deviations from uniformity become visible:
\begin{itemize}
\item[(i)] The retrogradation  loops and the angles $\beta$ of maximal elongation do not cover always the same arc but change with the position of the deferent in the ecliptic (so-called ``solar anomaly'').
\item[(ii)] 
 The mean motion (modelled by the   deferent)   deviates from uniform circular motion  also for  superior planets, similar to the sun's anomaly during the course of the year (``zodiacal anomaly''). 
\end{itemize}
 A change of the progression of synodic events (e.g. the forward stationary point) in the ecliptic was 
known to the Mesopotamian astronomers; they accounted for it by changing the arithmetical rules for the progression of two consecutive events in different parts of the  ecliptic.\footnote{The change was  implemented roughly by a step function in the Mesopotamian system $A$ and more subtly  by a piecewise linear  (zig-zag) function  in system $B$ \citep{Neugebauer:HAMA,Waerden:Astronomie},\citep[p. 42ff.]{Aaboe:Episodes2}.} 
In their way the Mesopotamian astronomers  thus knew of irregularities  corresponding to the effect (ii).
 In the epicyclic model arising from  kinematic  inversion of the heliocentric picture both  effects  can be taken into account only {\em separately}, and ad hoc,  by relaxing the condition that the terrestrial observer $T$ is placed at the center $C$ of the deferent circle, e.g., by an eccentric model like in figure \ref{fig exc deferent}. In any case the maximal error in longitude of an eccentric epicyclic model stayed than at the order of magnituide of $10^{\circ}$ for Mars.\footnote{\citep{Evans:1984,Rawlins:1987}.}
 
The geometry of kinematic inversion  is of course valid independently of  assuming uniform circular motion; an example has been given above for the simple eccentric model, fig. \ref{fig ecc S-T}. 
Applied to the Kepler kinematic of elliptical orbits it may help the modern reader to understand and to analyse  the ad hoc moves of ancient astronomers to bring their epicyclic models closer to the observed phenomena.  
An approximation of Kepler kinematics by a heliocentric equant model for the orbits of outer planets and a simplified uniform circular motion of the earth would, e.g.,  boil down to Ptolemy's equant model after the  kinematic inversion (see end of sec. \ref{section discussion}). 

	

\section{\small Another look at  pre-Ptolemaic epicyclic theory (historical argument) \label{section hist hypothesis}}

\subsection{\small Heliocentrism and early work with epicyclic models (Aristarchos to Seleukos) \label{subsection origin epicycles}}

The argument developed above strongly suggests  that the method of epicycles was born  during the 3rd century BC as a consequence of the researches of Greek astronomer-mathematicians who investigated the heliocentric hypothesis. As the sources on heliocentricism during the Hellenistic period are extremely sparse, we can neither be completely sure about this, nor ascribe this achievement to a specific person or school with any acceptable certainty.  
It is a pity that we have extremely thin documentary evidence on the work of Aristarchos who would  be an obvious candidate for such insights. In any case,  he is generally accepted as the first ``serious'' (i.e., mathematically literate) proponent of the heliocentric view in Hellenistic times. He  had good reasons for such a view  on the basis of his estimates for the comparative  sizes and distances of the earth, moon and sun, even though they are far away from later, more precise, estimates \citep{Aristarchos:Distances}.\footnote{His only transmitted work is \citep{Aristarchos:Distances}. The most reliable information of him having seriously defended the heliocentric hypothesis stems from Archimedes speaking about Aristarchos: ``His  hypotheses are that the fixed stars and the sun remain unmoved, that the earth revolves about the sun on the circumference of a circle, the sun lying in the middle of the orbit, and that the sphere of the fixed stars, situated about the same centre as the sun, is so great that the circle in which he supposes the earth to revolve bears such a proportion to the distance of the fixed stars as the centre of the sphere bears to its surface.'' \citep[p. 222 (409)]{Archimedes:Sandreckoner}. Another testimony for Aristarchos' helicentric view is Plutarchos {\em De facie in orbe Lunae}, quoted in \citep[p. 526]{Waerden:heliocentrism}.} 

We do know  that half a century later  Apollonios of Perga investigated  epicyclic kinematics.
According to Ptolemy, Apollonios studied the equivalence between a simple eccentric circular motion and an adapted concentric model with one epicycle ({\em Almagest} XII.1); moreover he  gave a geometric characterization of the stationary points of a retrogadation loop for an epicyclic kinematic model (of the type shown in fig. \ref{fig S-T-P},  right). 
He is thus often  considered as the {\em inventor} of the methods of epicycles.\footnote{For example in \citep{Kragh:Cosmos,Evans:1984,North:FontanaAstronomy,Pedersen:Almagest}.}
Such an attribution is highly   problematic, however,  because the  source gives  only evidence for him being an ``auctor  ante quem'', an author up to whom the method of epicycles has been introduced. Be this as it may,
  Ptolemy's report is clear evidence that  the method of epicyclic   models including  eccentric decentration  of circular motion was in use during  the late 3rd century BC already. Moreover it shows that the study of kinematic equivalence of different models between heavenly motions was a topic for mathematicians at that time.   

What was the motivation for such investigations? Ptolemy remained silent on this question, and Apollonios' work is not preserved. But for describing  the phenomenology of the  heliocentric point of view the  question analysed by Apollonios  makes perfect sense: 
The simplest method for producing a non-uniform motion of the apparent sun $S$ is to assume an eccentric dislocation of the true sun $S^{\ast}$ from the circle's center $A^{\ast}$. Then the kinematic inversion of the earth's true motion results in  a motion of the  apparent  sun $S_{\astrosun}$ on an eccentric circle with regard to the earth $T$, which deviates from the uniform motion of the mean sun $S_m$ (fig. \ref{fig ecc S-T}). As this is a disturbing feature,  one may  be inclined to ask whether the generated motion may be  substituted  by a construction using  the kinematic elements provided by the  first kinematic inversion, i.e. an epicyclic motion about the mean sun $S_m$. From this perspective, Apollonios's equivalence theorem appears no longer as a purely theoretical exercise. It also gives an answer to a  question arising naturally in an attempt for  bringing the kinematic inversion of the earth's motion into agreement with the observation of the sun's yearly course through the ecliptic.

His study of the stationary points of retrograde motions in an epicyclic model indicate a strong interest in the quantitative adaptation of the latter  to observational data. In the light of Archimedes' clear and respectful reference to Aristarchos it would seem  strange to assume that Apollonios, only a generation later,  did not know about the heliocentric hypothesis. And a geometer like him would surely be aware that the motion of  planets and the sun are given by epicyclic constructions  for a terrestrial observer.
We do not know whether Apollonios considered  the heliocentric hypothesis as realistic,  but a mathematical investigation of the latter could be done  independent of whether one shared the ontological perspective of Aristarchos. His investigation   of eccentric motions shows that Apollonios was well aware of the fruitfulness of considering kinematically equivalent constructions. It would be plausible to ascribe the detection of the second kinematic  inversion for the motion of outer planets to him.

It is difficult to specify when  Mesopotamian planetary tables became known to Greek astronomers, although it seems to have taken place during  3rd/2nd c. BC.\footnote{``Transmission of mathematical astronomy, chronologically, (\ldots) seems more or less with the transmission of astrology, probably not much before 200 B.C.''  \citep[p. 613ff.]{Neugebauer:HAMA}. } 
We know from Ptolemy that at this time observational planetary data were already taken by Greek astronomers with quite some precision  
({\em Timocharis}  in the year 272 BC, see above).
It may be that the Babylonian astronomical tables (ephemerides) were considered  a challenge to Greek astronomy  already at the time of Apollonios; at least it is clear that the problem of a quantitative adaption of the epicyclic or homocentric spherical  geometrical models   to more or less realistic data was posed and had to be attacked already from the 3rd century BC onward.

Thus the claim  that this challenge was {\em not}  accepted by the supporters of the heliocentric hypothesis  appears rather outlandish.  Just to the contrary, the close systematic connection between exploring the visible phenomena at the sky on the basis of a heliocentric assumption with the study of epicyclic constructions {\em and} the latters'  rise  during the 3rd century BC makes it highly plausible that the method of epicycles arose in the context of researches aiming at drawing not only qualitative, but also quantitative consequences from the heliocentric hypothesis. In contrast to what is usually stated in the history of Hellenistic astronomy (see sec. \ref{section discussion}) it {\em was possible} for Greek astronomers of the 3rd and 2nd centuries BC to adapt their kinematic models arising from the heliocentric picture to observational data (sec. \ref{subsection superior planets quantitative}) and to check their reliability with the the course of time. Due to the lack of sources we cannot, of course, prove without doubt that {\em they did so}.


 Plutarchos informs us about   another astronomer of Hellenistic time,   {Seleukos} of Seleukeia (Mesopotamia), $\sim$ 180 -- $\sim$ 120 BC, who  probably knew about  both, the Mesopot\-amian and the Greek traditions.  In his  remark, Plutarchos'  distinguishes between  Aristarchos who {\em supported the  hypothesis} (\textgreek{<upoti{j}'emenos m'onon}) of heliocentrism, while Seleukos   {\em  declared  it to be true}, or {\em demonstrated} it (\textgreek{ka`i >apofein'omenos}).\footnote{{\em Platonicae questiones} 1006C, here cited from \citep[p. 515]{Heath:Aristarchos}, compare \citep[p. 528]{Waerden:heliocentrism}.}  
It has  been discussed by different authors what this linguistic difference may mean, but without convincing results.\footnote{\citep{Waerden:heliocentrism,Russo:Revolution}.}
On the other hand it is clear  that Seleukos, writing a century after Aristarchos,  was in a much better position to draw  quantitative consequences of the heliocentric picture than the initiator of the hypothesis. It is possible that he was  able to compare some of its results with ephemeride predictions of the Mesopotamian systems $A$ or $B$. 
This presupposed more complicated chord-calculations than needed for the first quantitative determination of parameters discussed in section \ref{subsection superior planets quantitative}, at least for a series of selected values. 
But why should one not assume that the calculational  skills  necessary for the preparation and application of chord tables attributed to Hipparchos were  shared by other members of the community of his time?

 As long as we are lacking direct  sources on   heliocentrism   during the 3rd and 2nd centuries BC it is recommendable  to complement the indirect source evidence with  investigations of systematic   relationships between the heliocentric view and  the epicyclic method. This relationship is so close that we can state the following:
  \begin{itemize}
\item[--] Epicyclic constructions were a topic for mathematician-astronomers of the 3rd century, in particular Apollonios; but this does not signify that Apollonios  {\em invented}  the method of epicycles. 
 \item[--]  It is   likely that the epicyclic method arose as a result of  attempts   at translating  heliocentric planetary motions into the apparent kinematic seen from the earth. We do not know when this happened, but we can say that it took place between Aristarchos and Apollonios (boundaries included, with preference for the first one). 
\item[--]  In the 3rd and 2nd centuries BC Greek astronomy was characterized by two different competing geometric programs: the traditional homocentric sphere approach of Eudoxos and Kallypos, in which the central position of the earth was a fundamental assumption,  and a new one built on the method of epicyclic models which had its  background and origin in a kinematic inversion of the heliocentric perspective of Aristarchos.  In the latter case the central position of the earth in the epicyclic constructions played here only a perspectival, not an ontological fundamental role. 
 
 \item[--]  Eccentric deferent models (figure \ref{fig exc deferent}) were known and studied  already at the time of Apollonios. They were  explored as tools for generating deviations from uniform motion of the planets in the sky. Although standing in tension with an  ``orthodox''   heliocentric view of planetary motion, they may have been considered  for some while as an acceptable modification  of it. 
 \item[--]    Mesopotamian counts of long periods, enriched by observations of Greek astronomers of the 3rd and 2nd centuries BC, added by simple chord calculations sufficed for a first adaption of  epicyclic models to observational data, including a realist interpretation of the proportion of radii as relative distances of the planets from the sun.  
 It was thus possible to derive basic quantitative predictions from the heliocentric view of planetary motions.  It is possible that at the time of Seleukos more complex comparisons with the discrete time series for planetary ephemerides derived in the Mesopotamian systems were performed. 
 
 \end{itemize}

In his challenging book Russo pushed the thesis to the extreme that  Seleukos was convinced of  a dynamical underpinning of the heliocentric view by considerations of the contemporary research on the origin of tides  \citep[sec. 10.12]{Russo:Revolution}. I agree that it may very well be that dynamical  considerations played an important supporting role for the heliocentric perspective in the time between Aristarchos and Seleukos. But  the increasing importance of eccentric modifications of epicyclic models in astronomy must have contributed to undermining its plausibility. 
In case some supporter of heliocentrism had  played with the idea of, e.g.,  some  kind of ``counter-sun'' as a true center of the dynamics (in analogy to the ``counter-earth'' postulated by some Pythagorean thinkers in early Greek science) in order to make eccentric epicyclic models compatible with heliocentrism, the increasing precision knowledge about different types of eccentricities must have aggravated the problems for  the paradigm.

\subsection{\small Separation of epicycles  from a heliocentric background (Hipparchos to Ptolemy) \label{subsection Ptolemy}}
These problems became manifest in the work of Hipparchos in the middle of the 2nd century BC. 
 Closer inspection of planetary  data by Greek astronomers showed that for different superior planets the deferents have to be decentered differently, which was a problem for a dynamical interpretation of the heliocentric hypothesis. 
Even worse, with rising observational precision the existence of {\em two anomalies} for  the motion of superior planets became apparent:
	\begin{itemize}
	\item[--]   uneven motion of $D_2$ in the ecliptic (in later terms, due to Kepler's second law)
	\item[--]  	and uneven retrograde arcs for $Ps$ (due to Kepler's first law)
\end{itemize}
	The two  effects were {not reducible}  to one eccentric mounting of the deferent. 
	 Ptolemy describes the critical evaluation of the planetary theories of the 2nd century BC by Hipparchos   as follows ({\em Almagest}. IX.2):

\begin{quote}
\ldots he [Hipparchos] thought that one must not only show that each planet has a twofold anomaly, or that each planet has retrograde arcs which are not constant, and are of such and such sizes (whereas the other astronomers had constructed their geometrical
proofs on the basis of a single unvarying anomaly and retrograde arc); {nor} that these anomalies can in fact be represented {either by means of eccentric circles or  by circles concentric with the ecliptic and carrying epicycles}, or even by combining both, the ecliptic anomaly being of such and such a size, and the synodic anomaly of such and such \ldots 
\citep[p. 421, translation Toomer]{Ptolemaios/Toomer:Almagest}
\end{quote}
This is clear evidence of the {\em existence}  of quantitative epicyclic models with simple eccentricity in the 2nd c. BC. and their {\em invalidation} by  Hipparchos on the basis of observational data most of which were inherited from the Mesopotamian astronomers. Ptolemy's text continues:
\begin{quote}
\ldots (for these
representations have been employed by almost all those who tried to exhibit the
uniform circular motion by means of the so-called `Aeon-tables', but their
attempts were faulty and at the same time lacked proofs: some of them did not
achieve their object at all, the others only to a limited extent); \ldots (ibid., p. 122)
\end{quote} 
This is an  illuminating remark. It states explicitly that in the time of Hipparchos the quantitative evaluation of epicyclic models was not restricted to   sporadic, specifically selected data of interest, but also to the preparation of {\em tables of long duration}.\footnote{``Aeon-tables'' in Toomer's translation, ``perpetual tables'' in \citep[p. 789]{Neugebauer:HAMA}, ``Tafeln f\"ur ewige Zeiten'' in the German translation by K. Manitius, 1963.}

Even if  epicyclic models were initially motivated by  the heliocentric hypothesis, once they were introduced they could be used  {pragmatically} without subscribing to the  commitment to heliocentrism. This was apparently  so for Hipparchos, an astronomer-mathematician in contrast to Aristarchos and Apollonios who were mathematician-astronomers.
It is well known that he developed  numerical methods for astronomy in terms of  Mesopotamian sexagesimal numbers. With such methods he established a successful moon model and a  sun model, while he did not propose an alternative model of his own for the planets (in the present sense) \citep{North:FontanaAstronomy}. 

	After  the invalidation of simple eccentric deferent models by   Hipparchos,  perhaps even already before it,  ``serious'' astronomers may have looked for a better  fit of epicyclic models to the observational data without bothering about ontological questions, a late Hellenistic version of {\em scientific positivism}. 
The  remark by Ptolemy, quoted above, about the combination of eccentric deferents (for one anomaly) and an additional epicycle (for the other anomaly) seems to hint in this direction. 
A further influx of Mesopotamian  data and calculational methods, accompanying astrological practices  during the 2nd/1st centuries  BC  \citep[pp. 608, 613]{Neugebauer:HAMA} must have led to increasing difficulties for epicyclic kinematics, and with it for heliocentrism. 
In the time between  Hipparchos and Ptolemy (roughly the 1st centuries BC and AD) a diversity of kinematic  {ad hoc models}  for different planets were developed.\footnote{For the study of and  the confusion about  how to deal with the ``two anomalies'' of planets one may consult \citep[801ff.]{Neugebauer:HAMA}. Additional historical evidence for crude and indecisive  modifications of eccentric models with different perigees is given in  Plinius \emph{Historia Naturalis}, book II sec. xvi.}

\begin{figure}

\hspace*{8em}\includegraphics[scale=0.7,angle=0]{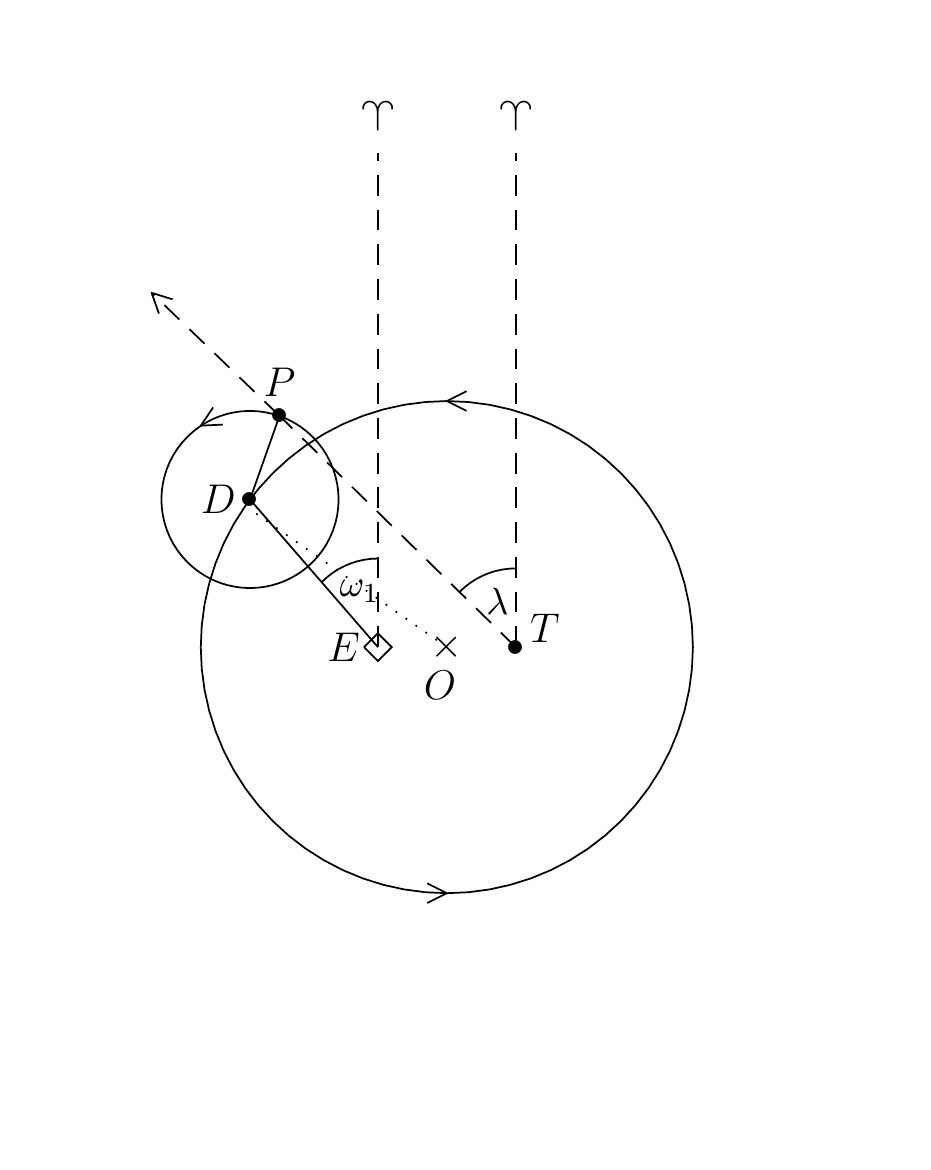}
\caption{Epicyclic motion of $P$ with  eccentric deferent $D$  and equant $E$. $D$ moves non-uniformly on a circle about $O$ decentered from the earth $T$; but the angle $\omega_1= \angle \aries E D$ progresses uniformly (constant angular velocity). In principle  $E, O, T$ need not  be collinear, but by empirical reasons (for ancient astronomers), they are (even with $\overline{OE}=\overline{OT}$). $\lambda(t)$ is the ecliptic  longitude angle of the planet for an observer at T.  \label{fig equant}}
\end{figure}

A  solution of the double anomaly problem for epicyclic kinematics was finally found by introducing a separate point $E$ relative to which the motion of the deferent $D$  on a circle with center  $O$ appears uniform, while the observations is made from another decentered point $T$ (earth). The center of uniformity $E$ is called the {\em equant}  (figure \ref{fig equant}) (\emph{Almagest}, IX.6ff).
 Then   two different  perigees arise: direction of $TO$ (about which the smallest retrogradation arc is observed)   and the direction  of $EO$ in which the  angular velocity of the deferent is smallest. 
Observational evidence of Mars may have led to a collinear placement of $E, O, T$   and  $O$ in the midpoint of $ET$.\footnote{A systematic reconstruction is given in  \citep{Evans:1984}.}
 This led to a satisfying kinematic model for the outer planets. The  historical origin of the equant construction is unclear; most authors praise  Ptolemy for it, but there are  indications of an earlier use of the method of an  eccentric deferent circle with equant  \citep {Duke:2005India} and before him \citep{Waerden:1961Ausgleichspunkt}. 
 Observations of the strong ``solar anomaly'' for Mars (varying retrograde arcs) seem  to predate Ptolemy considerably, as is the case for large parts of the observational data collected in his famous fixed star catalogue.\footnote{See the detailed historical study \citep{Grasshoff:1990}, a short description in \citep{Grasshoff:2014}.} 

We do not know when and by whom the equant method was invented;  so we  cannot exclude that it may have  been designed  originally in a  heliocentric framework for  modelling the non-uniform ``true'' motions of the planets (in particular for Mars) and was  translated by kinematic inversion to the geocentric model later published by Ptolemy.\footnote{Rawlings even assumes that this was the case not only for the orbit of the outer planets but for all of them including the earth. This would imply not only an equant construction for the deferent, but  also a decentered epicycle with equant, surpassing the precision of  Ptolemy's model by far \citep{Rawlins:1987}. This appears quite far-fetched; Rawlins' conjecture seems to be the result of an ex-post approximation of the  Kepler dynamics by decentered circular orbits about the sun with symmetric equant. The historical arguments given in the paper are quite meager.}
If so, such researches  probably have added to the irritations for the ontological hypothesis of heliocentrism. Moreover,  the determination  of parameters and the check of empirical adequacy could be done just as well, and from the pragmatic point of view even more directly in the geocentric version of the equant approach. For practising astronomers without philosophical ambition  the heliocentric origin of the approach may  have appeared more and more as ballast which could just as well be dropped, as long as dynamical explanations in mathematical form were out of reach.\footnote{Russo disagrees with this estimation, see end of sec. \ref{subsection origin epicycles}, and so does V. Blasj{\o} (personal communication).}

 With the development of calculational tools the question of heliocentrism versus geocentrism lost its practical importance even more because of the kinematic equivalence of the pictures.\footnote{For early medieval  Indian astronomy, in particular for Aryabatha (5th c. AD) it was apparently a superfluous question \citep{Subramanian:1994,Subramanian:1998},  \citep[p. 111ff]{Plofker:2009}. This does not  contradict van der Waerden's interpretation of Aryabatha's identification of the apogees of Venus  and the true sun  as a {\em trace} of a {\em heliocentric} origin \citep[p. 532]{Waerden:heliocentrism}. This view fits well to the evidence of pre-Ptolemaic Hellenistic influence on Aryabatha's astronomical scheme collected by  \citep{Duke:2005India}.}
For Copernicus, however,  the equant approach became a stumbling stone which he had to overcome for his attempt to reverse  the second kinematic inversion and to reconstruct a heliocentric picture as he understood it. But his  leads over to a different story, namely the replacement of the equant by additional epicycles and the long story leading finally to Kepler's detection of elliptic kinematics. 

\pagebreak
\section{\small  Discussion and final remarks \label{section discussion}}
In large parts of the literature on pre-Ptolemaic astronomy  the heliocentric approach of Aristarchos and successors is attributed only a marginal role. It seems to be a  widely shared consensus that ancient heliocentrism was a speculative play of ideas only,  without a noticeable effect on the course of Greek astronomy. 
 Neugebauer argues: 
``Without the accumulation of a vast
store of empirical data and without a serious methodology for their analysis the
idea of heliocentricity was only a useless play on words'' \citep[p. 698]{Neugebauer:HAMA}.
In a similar vain H. Kragh sees a contrast between the ideas of a mathematician (Aristarchos) against the more serious and  empirically founded investigations of astronomers like Hipparchos and Ptolemy \citep[p. 27]{Kragh:Cosmos}. Exceptions of this consensus view come from  outsiders only, like \citep{Waerden:heliocentrism,Rawlins:1987} and more recently \citep{Russo:Revolution}.

The argument presented here is that any mathematician-astronomer who is interested in observations and has only classical geometrical tools at his disposal cannot but  introduce epicyclic constructions. This argument is so simple that it should have been considered in the  literature long since. The lack of sources is surely a problem, but we have similar effects in other parts of ancient history of science, e.g. in the field of pre-Euclidean mathematics. Nobody would claim that demonstrative mathematics started only with Euclid (and perhaps Eudoxos), because pre-Euclidean or pre-Eudoxian sources on Greek mathematics do not exist and its contributions  can be reconstructed only tentatively from later fragmentary reports and indirect traces in the transmitted corpus of knowledge. 

In the case of those parts of pre-Ptolemaian astronomy which were not taken up in the later main corpus of texts 
 this seems to be handled differently. We  are even confronted with  trivially circular arguments of the form:
\begin{quote}
It seems unlikely that Aristarchus developed any specific planetary theory because  our sources are silent on this topic; \ldots \citep[p. 692]{Neugebauer:HAMA}
\end{quote}
From the perspective developed in the present paper this  appears highly surprising. It  seems much more likely that planetary theories based on the  heliocentric view were formulated and 
had a long range influence through their contributions to the epicyclic method. But why was this no longer explicitly acknowledged in the later source literature? Probably this was the result of two mutually enforcing causes. One was the increasing irritation of the complicated ad-hoc modifications of epicyclic models necessary to cope  with the increasing precision of the empirical knowledge on planetary motion. This   has been discussed above (sec. \ref{section hist hypothesis}). The other was apparently of  a more ideological nature: Heliocentrism was considered as a  heterodox and even heretic  perspective. This seems to have led to the exclusion of an open acknowledgement of heliocentric convictions or influences.\footnote{In \citep{Russo:Revolution} the cultural degeneration in the Roman Imperial period of post-Hellenistic antiquity, and ideological factor are seen as  the predominant, if not the only cause.} 
 
 Once our eyes are opened, we  may start to  see a lot of traces of heliocentrism in the literature on the epicyclic method like, e.g., the grouping of the planets into inferior and  superior ones and their succession, the role of the mean sun in epicyclic constructions, the realistic interpretation of distances in the epicyclic picture etc. We cannot judge, whether Ptolemy still knew about the heliocentric background to the epicyclic method and preferred not to write about it, or whether at his time (and in his position) the knowledge was already suppressed.  But some of his remarks  may indicate at least a remnant of the recollection of a more systematic origin of the epicyclic method than he could, or wanted to, explain. In the passage  cited above on principles of planetary motions in which he discussed Hipparchos' contributions he continued:
 \begin{quote}
 The point of the above remarks was not to boast. Rather, if we are at any point compelled by the nature of our subject to use a procedure not in strict accordance with theory (\ldots); or [\ldots] to make some basic assumptions which we arrived at not from some readily apparent principle, but from  a long period of trial and application, \ldots
we know too that assumptions made without proof, provided only that they are found to be in agreement with the phenomena, {\em could not have been found without some careful methodological procedure; even if it is difficult to explain how one came to conceive them} (\ldots) we know, finally, that some variety in the type of hypotheses associated with the circles [\ldots] cannot plausibly be considered strange or contrary to reason; \ldots \citep[p. 395f., emph. ES]{Ptolemaios/Toomer:Almagest} 
 \end{quote}
This is a cryptic remark hinting   vaguely at some {\em methodological rationality  behind} the  principles or assumptions  which were  often introduced by  Ptolemy relatively ad hoc for his final model. It is open to interpretation. I tend  to interpret it as a possible indirect hint at a heliocentric rationality behind the construction of the final epicyclic model of planetary motion in the {\em Almagest}. 

Of course, all this deserves more historical scrutiny. But a precondition for this is an  acceptance at least of the possibility, if not of a high plausibility, that  heliocentrism has had the potential to act as an effective component in the rise of epicyclic models, i.e., in a crucial phase of Greek astronomy.\\[3em]

\subsubsection*{Acknowledgements:} The interpretation presented in this paper arose in the discussions during a joint Wuppertal seminar on ancient astronomy  with Friedrich Steinle (summer  2009), which  I remember  with joy. I am grateful for helpful comments 
by Helge Kragh and Victor Blasj{\o}.  The figures have been drawn with Klaus Fritzsche's  graphics package mga-21.\footnote{\url{http://www2.math.uni-wuppertal.de/~fritzsch/kf_latx.html}}


\addcontentsline{toc}{section}{\protect\numberline{}Bibliography}
\small
 \bibliographystyle{apsr}
 \bibliography{%
/home/erhard/Dropbox/Datenbanken/BibTex_files/lit_hist,%
/home/erhard/Dropbox/Datenbanken/BibTex_files/lit_mathsci,%
/home/erhard/Dropbox/Datenbanken/BibTex_files/lit_EB-LAIIIB,%
/home/erhard/Dropbox/Datenbanken/BibTex_files/lit_scholz}


\end{document}